\documentclass[10pt,twocolumn]{article}
\usepackage[top=0.8in,bottom=0.8in,left=0.65in,right=0.65in,columnsep=0.25in]{geometry}
\usepackage{mathptmx}
\usepackage[T1]{fontenc}
\usepackage{amsmath,graphicx,lineno,setspace,caption,booktabs,array}
\usepackage[numbers,sort&compress,super]{natbib}
\usepackage[hidelinks]{hyperref}
\newcommand{\kms}{km\,s$^{-1}$}
\newcommand{\Ha}{H$\alpha$}
\newcommand{\Hb}{H$\beta$}
\newcommand{\oiii}{[O\,\textsc{iii}]}
\newcommand{\vesc}{$v_{\rm esc}$}
\renewcommand{\refname}{References}

\usepackage[scaled=0.92]{helvet}
\usepackage{titlesec}
\titleformat*{\section}{\large\bfseries\sffamily}
\titleformat*{\subsection}{\normalsize\bfseries\sffamily}
\titlespacing*{\section}{0pt}{8pt}{3pt}
\titlespacing*{\subsection}{0pt}{6pt}{2pt}
\begin{document}

\twocolumn[{\begin{center}
{\LARGE\bf\sffamily Black holes in little red and blue dots drive no superwinds in the first 1.5 billion years}\\[10pt]
Mainak Singha$^{1,2}$, Yurina Nakazato$^{3}$ and Wei Leong Tee$^{4,5}$\\[6pt]
{\small $^{1}$Astrophysics Science Division, NASA Goddard Space Flight Center, Greenbelt, MD, USA. $^{2}$Department of Physics, The Catholic University of America, Washington, DC, USA. $^{3}$Center for Computational Astrophysics, Flatiron Institute, New York, NY, USA. $^{4}$Department of Astronomy and Astrophysics, The Pennsylvania State University, University Park, PA, USA. $^{5}$Institute for Gravitation and the Cosmos, The Pennsylvania State University, University Park, PA, USA.}
\end{center}

{\bfseries Galaxy-formation models rely on accreting black holes to expel gas from their hosts. We test this prediction in 67 little red dots (LRDs) and little blue dots (LBDs) in the first 1.5 billion years, selected from 80{,}367 NIRSpec spectra in the DAWN JWST Archive. We detect no \oiii\ wind, against lower limits of 7.2 and 7.5 expected detections for local sub-Eddington and super-Eddington active nuclei. For winds narrower than 1{,}500 \kms, the corresponding incidence limits are 13\% and 19\% at 95\% confidence, compared with 31\% and 49\% locally. Broad Balmer lines are centred on the narrow lines, and Balmer absorbers move below 8\% of the escape speed. Published super-Eddington fountain, black-hole-star and quasi-star models each fail a measured property of the spectra. LRDs and LBDs drive no superwinds, and their black holes grow without the wind feedback invoked to regulate galaxies.\par}

\vspace{14pt}}]

How do growing black holes begin to regulate their host galaxies? Galaxy-formation models invoke black-hole feedback to limit galaxy masses and shut down star formation\cite{benson03,bower06}. Superwinds expel galactic gas\cite{heckman90}; AGN feedback models invoke radiation-driven winds and jet heating\cite{fabian12}. Near the Eddington limit, radiation drives superwinds beyond the escape speed of the host\cite{silk98,dimatteo05}; at lower accretion rates, jets heat the surrounding halo\cite{bower06}. In the model of Silk and Rees\cite{silk98}, a quasar-driven shell escapes at about 500 \kms, and the balance between the wind and gravity sets the relation between black-hole mass and stellar velocity dispersion. Rapid black-hole growth is thus tied to strong feedback.

The little red dots discovered by JWST\cite{matthee24,greene24,kocevski25,hviding25} challenge this connection during the first 1.5 billion years. Their virial black-hole masses lie above the local relation between black-hole and stellar mass\cite{pacucci23}, yet the sources are weak in X-rays\cite{ananna24,yue24,maiolino25} and at radio wavelengths\cite{mazzolari24,akins25}, and most lack the high-ionization ultraviolet lines characteristic of local active galactic nuclei (AGN)\cite{lambrides24}. Super-Eddington models explain these properties through intrinsic X-ray weakness, broad lines that overestimate the black-hole mass, and radiation reprocessed by dense gas\cite{pacucci24,lambrides24,naidu25,kaaz26}. They also predict powerful winds. Radiation-magnetohydrodynamic simulations of super-Eddington discs launch outflows carrying about 30\% of the accreted mass\cite{jiang14,sadowski16}, at $10^{3}$--$10^{4}$ \kms\ or more\cite{kaaz26}. A dormant overmassive black hole\cite{juodzbalis24} and a dynamical black-hole mass above the virial estimate\cite{juodzbalis26} have, however, motivated alternatives to super-Eddington growth.

Wind kinematics test these interpretations directly. NIRSpec spectra probe ionized gas in the host, the broad-line region and absorbing gas along the sightline to the nucleus. An ionized outflow can produce a blueshifted \oiii\ wing, whether its emission is compact or extended\cite{heckman81,singha22,winkel23}; a broad-line-region wind can shift or skew the Balmer emission; and approaching foreground gas produces blueshifted Balmer absorption. These three diagnostics test the host gas, the broad-line region as a whole and individual nuclear sightlines. Ionized winds also occur in some weakly accreting local radio galaxies\cite{singha21}. Nearby type 1 AGN show \oiii\ winds in 31\% of sub-Eddington and 49\% of super-Eddington sources (Methods). We apply the same statistical selection to the primary spectra, injections and local comparison. From 80{,}367 NIRSpec spectra in the DAWN JWST Archive, we select 334 LRD and 1{,}927 LBD targets; 95 LRDs and 52 LBDs enter the broad-line analysis sample, and 67 have grating \oiii\ spectra deep enough for the wind search (Table~1). We first establish whether local-AGN-like winds would be detected, then use the nuclear kinematics and published model predictions to interpret their absence.

\begin{table*}[t]\centering\small
\caption*{\textbf{Table 1 $|$ Sample selection and the sources behind each test.} Upper block: selection steps (Methods, Sample). Middle block: sources entering each test. Lower block: local comparison samples. LRDs have V-shaped PRISM continua or are published LRD candidates; LBDs are compact sources outside the V-shaped selection. Parentheses give marginal broad-line detections in the selection block and subsets in the test blocks.}
\begin{tabular}{>{\raggedright\arraybackslash}p{9.6cm}rr}\toprule
 & LRDs & LBDs\\\midrule\midrule
NIRSpec spectra in the DAWN JWST Archive v4.4 & \multicolumn{2}{r}{80{,}367}\\
Targets with archive spectra & 334 & 1{,}927\\
Broad Balmer lines detected (marginal) & 99 (20) & 47 (32)\\
Analysis sample & 95 & 52\\\midrule
\oiii\ winds (Fig.~1, injections): usable grating \oiii\ spectra & 53 & 14\\
Broad \Ha\ centroids and extra components (Fig.~2, right): signal-to-noise ratio above 5 & 47 & 7\\
Broad \Ha\ stacked profiles (Fig.~2, left): signal-to-noise ratio above 20 & 30 & 3\\
Balmer absorbers (Fig.~3): detected (with \Hb\ coverage) & 14 (12) & 1 (1)\\
Fountain comparison (Fig.~4): \Ha\ wing deficit measured (best measured) & 35 (7) & 5 (3)\\
Envelope tests (Fig.~5): blackbody fit to the red continuum (broad \Ha\ signal-to-noise ratio above 40) & 68 (12) & --\\\midrule
 & \multicolumn{2}{c}{Local AGN}\\\midrule\midrule
\oiii\ winds: sub-Eddington (with winds) & \multicolumn{2}{r}{9{,}270 (2{,}908)}\\
\oiii\ winds: super-Eddington (with winds) & \multicolumn{2}{r}{412 (203)}\\
Broad \Ha\ extra components: SDSS, both ends of the quasar main sequence & \multicolumn{2}{r}{601}\\
Broad \Ha\ stacked profile: strong-Fe\,\textsc{ii} subset (Fig.~2) & \multicolumn{2}{r}{325}\\
Balmer absorbers (Fig.~3) & \multicolumn{2}{r}{6}\\
\bottomrule\end{tabular}\end{table*}

\section*{Results}

\subsection*{No \oiii\ winds in red or blue dots}

\begin{figure}[t]\centering\includegraphics[width=\columnwidth]{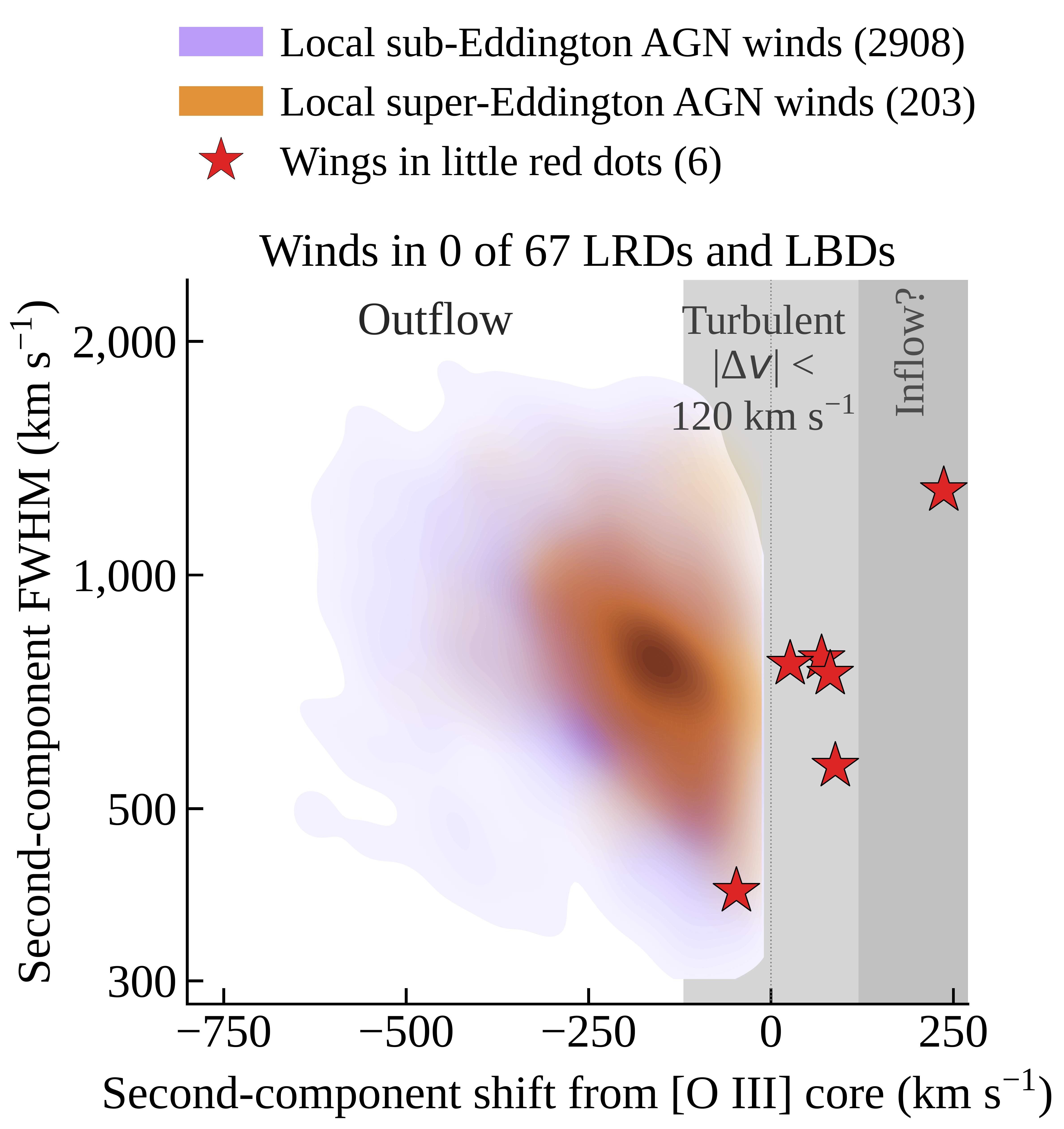}
\caption*{\textbf{Fig. 1 $|$ None of the 67 LRDs and LBDs shows a wind where local AGN winds lie.} Shift of the second \oiii\ component from the narrow core against its FWHM, for 2,908 local sub-Eddington winds, 203 local super-Eddington winds and the six significant second components in our sample (stars, all LRDs). The central band marks shifts below 120 \kms, the range of turbulent motions in ionized gas at these redshifts\cite{danhaive25}. Shaded regions play no role in detection.}\end{figure}

The primary search detects no \oiii\ wind candidates among 67 LRDs and LBDs (Fig.~1). We fitted one grating spectrum per source with and without a second \oiii\ component, allowing its width and velocity to vary. A candidate must pass the component-significance cuts and have a blueshift exceeding three times its velocity uncertainty (Methods). Six spectra, all of LRDs, require a second component. Five of these components lie within 120 \kms\ of the narrow core, within the range of turbulent ionized-gas motions at these redshifts\cite{danhaive25}: one at $-47\pm19$ \kms\ and four redshifted by 26--88 \kms. The sixth component lies at $+240\pm130$ \kms, a 1.8-standard-deviation redshift with no blueshifted counterpart. None meets the wind criterion (Methods).

Two of the 67 sources also have higher-resolution G395H spectra (Methods). Refits of these spectra find a second \oiii\ component in each, at $-72\pm12$ and $-38\pm9$ \kms, both blueshifted by more than three standard deviations. Neither refit replaces a primary fit or enters the 67-source count and injection--recovery statistics. Both shifts are below the host velocity dispersions inferred from the narrow lines and far below the escape speed, so neither component is a superwind.

Local AGN set the empirical expectation (Methods). We applied the same wind criterion to 9{,}682 local type 1 AGN: 208 from the BAT AGN Spectroscopic Survey and 22 from the Sloan Digital Sky Survey (SDSS), which we fitted with the same model, and 9{,}452 from the ALPAKA catalogue of SDSS AGN, whose published two-component \oiii\ decompositions we classified\cite{koss17,alpaka}. Dividing them at the Eddington limit, with bolometric luminosities from the narrow \oiii\ line, winds occur in 2{,}908 of 9{,}270 sub-Eddington and 203 of 412 super-Eddington AGN. Of these winds, 97\% have \oiii\ full widths at half maximum (FWHM) below 1{,}500 \kms. Their shifts span 200--1{,}200 \kms, encompassing the 500 \kms\ breakout speed of a quasar-driven shell\cite{silk98}.

We tested whether these winds would be visible by injecting artificial components spanning the widths, shifts and strengths of local winds into each of the 67 primary spectra and repeating the search. Combining the source-specific recovery probabilities with the local wind fractions gives lower limits of 7.2 expected detections for a sub-Eddington-like population and 7.5 for a super-Eddington-like population. The probabilities of detecting none are $4\times10^{-4}$ and $3\times10^{-4}$. Local wind incidence depends on black-hole mass and luminosity, so we also reweighted the local AGN to the distributions of our sources (Extended Data Fig.~2). Matching in mass and luminosity gives 8.0 expected sub-Eddington detections, and matching in mass gives 5.8 super-Eddington detections. Deriving the local bolometric luminosities from the broad Balmer lines, as for our sources, and matching in Eddington ratio gives 8.2. Every version leaves the chance of detecting none at $2\times10^{-3}$ or less (Methods). These expectations are lower limits because local winds broader than 1{,}500 \kms\ or slower than 200 \kms\ were counted as undetectable.

For winds narrower than 1{,}500 \kms\ drawn from the local sub-Eddington population, the 95\% confidence upper limit on incidence is 13\%, compared with 31\% locally. For the same width range and the local super-Eddington distribution, the limit is 19\%, compared with 49\% locally. These are two reference populations for a single 67-source test. If winds are intermittent and the sources sample random activity phases, the limits also constrain their duty cycle. The wind deficit persists against either local accretion regime.

\subsection*{The gas near the black hole is bound}

\begin{figure*}[t]\centering\includegraphics[width=\textwidth]{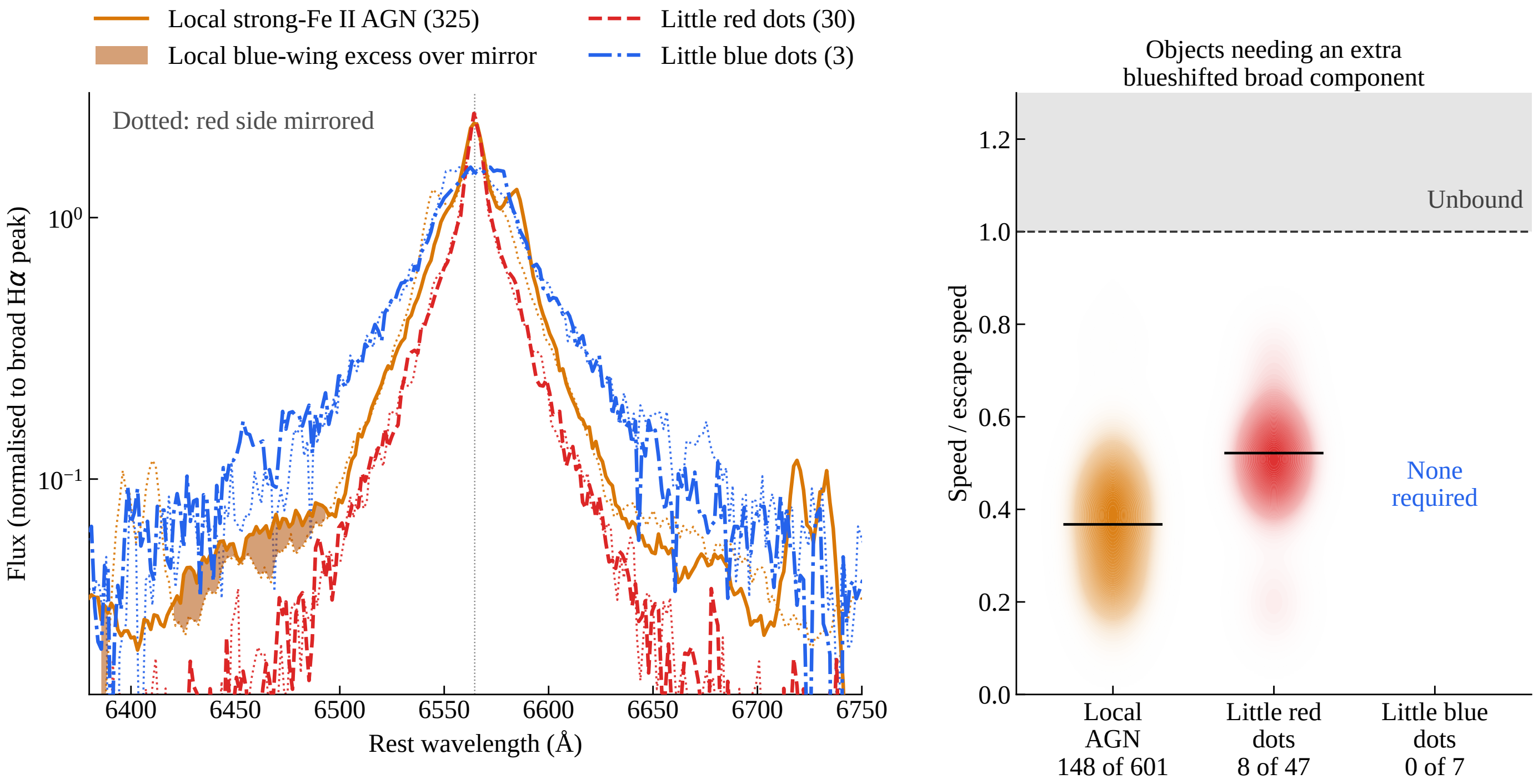}
\caption*{\textbf{Fig. 2 $|$ Broad Balmer lines in LRDs show no blue excess and stay bound.} Left: stacked broad \Ha\ profiles of 325 local strong-Fe\,\textsc{ii} broad-line AGN (solid), 30 LRDs (dashed) and three LBDs (dash-dotted), the sources with the highest broad \Ha\ signal-to-noise ratio (above 40 locally, above 20 in LRDs and LBDs; Table~1). Dotted curves of the same colour mirror each red wing about the line centre; the shaded band, drawn for the local stack only, marks its blue-wing excess over the mirror. Right: speed of the additional blueshifted broad component divided by the escape speed at the broad-line radius, for all sources with broad \Ha\ signal-to-noise ratio above 5. The counts give the sources requiring this component: 148 of 601 local AGN, 8 of 47 LRDs and none of seven LBDs. Bars indicate medians. Two local components with ratios of 1.3 and 1.6 lie outside the plotted range.}\end{figure*}
\begin{figure*}[t]\centering\includegraphics[width=\textwidth]{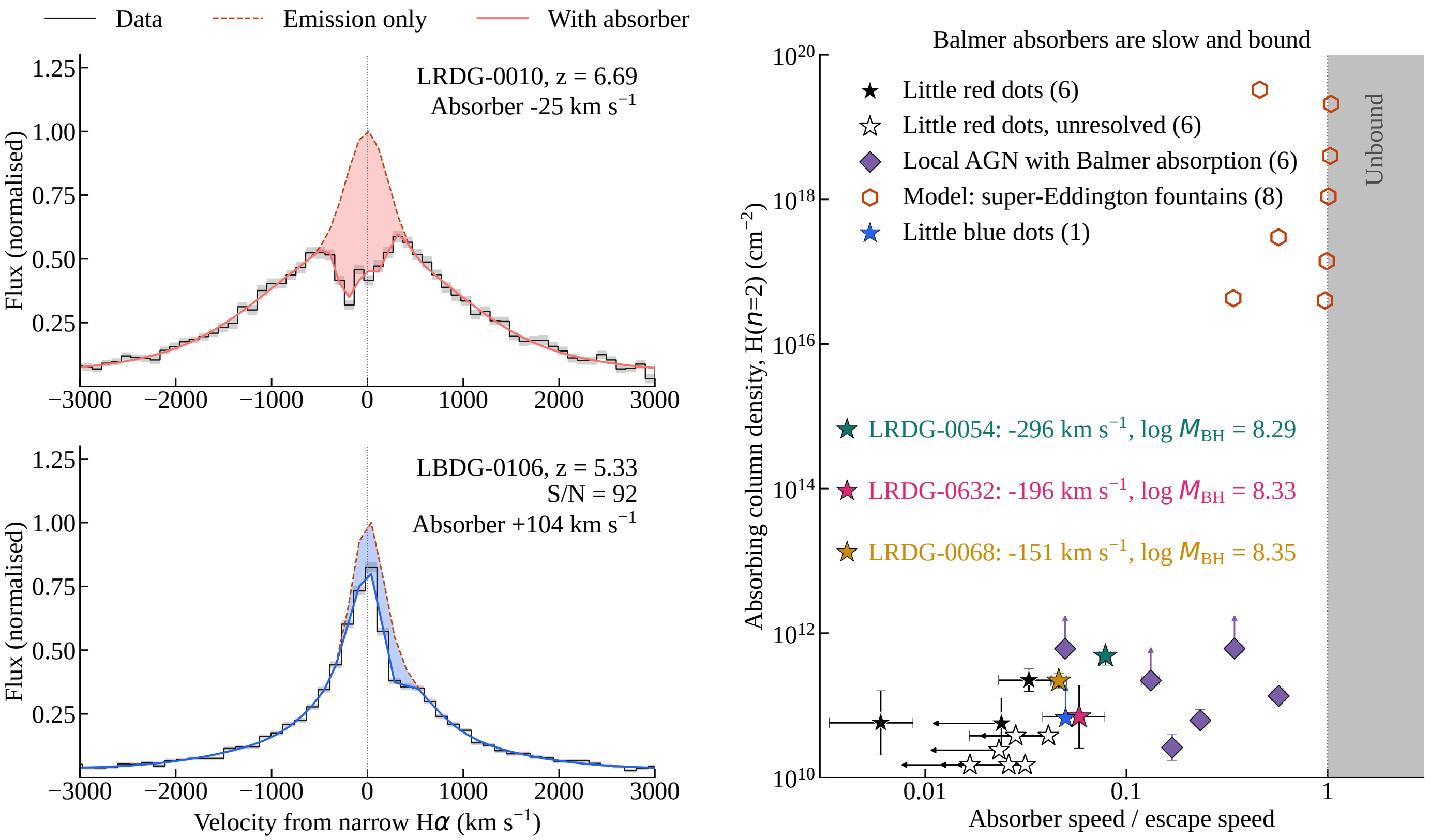}
\caption*{\textbf{Fig. 3 $|$ Balmer absorbers sit close to rest, far below the escape speed.} Upper left: \Ha\ and its absorption component in LRDG-0010. Lower left: the same for LBDG-0106. Right: hydrogen column density in the $n=2$ level versus absorber speed divided by the escape speed of the broad-line gas. Coloured stars mark the three fastest LRD absorbers, labelled with their velocity and black-hole mass. Measurements for LRDs, the LBD and local AGN are compared with super-Eddington fountain models (open hexagons)\cite{kaaz26}. Open stars along the bottom indicate six LRD absorbers narrower than the instrumental resolution: their velocities are measured, but their column densities are unconstrained. Left-pointing arrows denote two-standard-deviation upper limits for speeds consistent with zero.}\end{figure*}

Broad Balmer lines test whether winds are launched on smaller scales. Pacucci and Narayan\cite{pacucci24} proposed that strong winds enhance the broad lines of super-Eddington LRDs, causing virial estimates to overstate their black-hole masses. Such winds shift or skew the emission and accelerate gas along the sightline. Instead, broad \Ha\ is centred close to the narrow lines: the median offset is $+0.9\%$ of the line width in LRDs, compared with $-1.1\%$ and $+0.5\%$ in the two local samples. The local stacked profile has a blue-wing excess; the LRD and LBD stacks are symmetric (Fig.~2).

We searched individual broad lines for a blueshifted excess by adding a Gaussian to an exponential electron-scattering profile, then repeated the test with a redshifted Gaussian as a control. The test uses every source with broad \Ha\ signal-to-noise ratio above 5, namely 47 LRDs and 7 LBDs, and 601 SDSS broad-line AGN with signal-to-noise ratio above 40 that span both ends of the quasar main sequence (Methods). A blueshifted component is required in 8 of 47 LRDs, 148 of 601 local AGN and none of 7 LBDs: comparable fractions in LRDs and local AGN, 17\% versus 25\%, or 22\% after matching the local AGN to the black-hole masses of the LRDs (Fisher $p=0.57$). All eight LRD components move at 0.2--0.7 of the escape speed from the black hole at the broad-line radius (Methods); 146 of 148 local components also lie below this speed. A single scattering profile cannot follow an asymmetric line core. With two independent scattering profiles as the base, only 2 LRDs require an extra blueshifted Gaussian and 3 a redshifted one, compared with 36 and 50 in local AGN and none in LBDs; a two-Gaussian base gives the same result (Methods). Blue and red components occur in similar numbers: the additional components show no systematic preference for approaching gas.

Balmer absorption independently probes gas along the nuclear sightline. We detect absorbers in 14 LRDs and one LBD (Table~1). Fig.~3 shows a resolved LRD absorber near the median velocity, LRDG-0010, and the single LBD absorber, LBDG-0106, alongside the speeds and $n=2$ hydrogen column densities of all absorbers. The column densities follow from the optical depth and width of each trough (Methods). Joint \Ha\ and \Hb\ fits to the 12 LRD absorbers with \Hb\ coverage give velocities of $-296$ to $+9$ \kms\ relative to the narrow lines, with a median of $-37$ \kms; the other two have \Ha\ only and lie at $-81$ and $+26$ \kms. Six of the 12 are consistent with zero and three are blueshifted by more than three standard deviations. Their speeds are only 0.001--0.08 of $v_{\rm esc}$, the escape speed from the black hole at the broad-line radius, which we take equal to the broad-line FWHM (Methods); the LBD absorber moves at 0.05\,\vesc.

At broad-line-region radii, these speeds keep the gas bound. Absorption from the $n=2$ level of hydrogen requires densities of at least $10^{8}$ cm$^{-3}$ in the best-studied case\cite{juodzbalis24b} and above $10^{9}$ cm$^{-3}$ in models of LRD absorbers\cite{inayoshi25}, placing the absorbing gas close to the broad-line region. For ballistic radial motion, gas launched from radius $r$ at speed $v$ reaches $r_{\rm max}=r/(1-v^2/v_{\rm esc}^2)$ before falling back. At the fastest measured speed, $0.08\,v_{\rm esc}$, it travels only 0.6\% beyond its starting radius. Escape at that speed requires a launch radius beyond about 150 broad-line radii, far outside the region where $n=2$ absorption arises. In the super-Eddington fountains of Kaaz et al.\cite{kaaz26}, a slow disc wind partly escapes and partly falls back; its absorbing gas has column-weighted speeds of 0.34--1.04\,\vesc\ (Methods), compared with 0.05--0.57\,\vesc\ in nearby AGN with Balmer absorption. Every LRD and LBD absorber lies below the fountain-model range. The gas near the black hole is bound: on a ballistic orbit, the absorbing material rises less than 1\% beyond its launch radius, which suggests that it falls back rather than escapes.

\subsection*{Super-Eddington models fail the data}

\begin{figure}[t]\centering\includegraphics[width=\columnwidth]{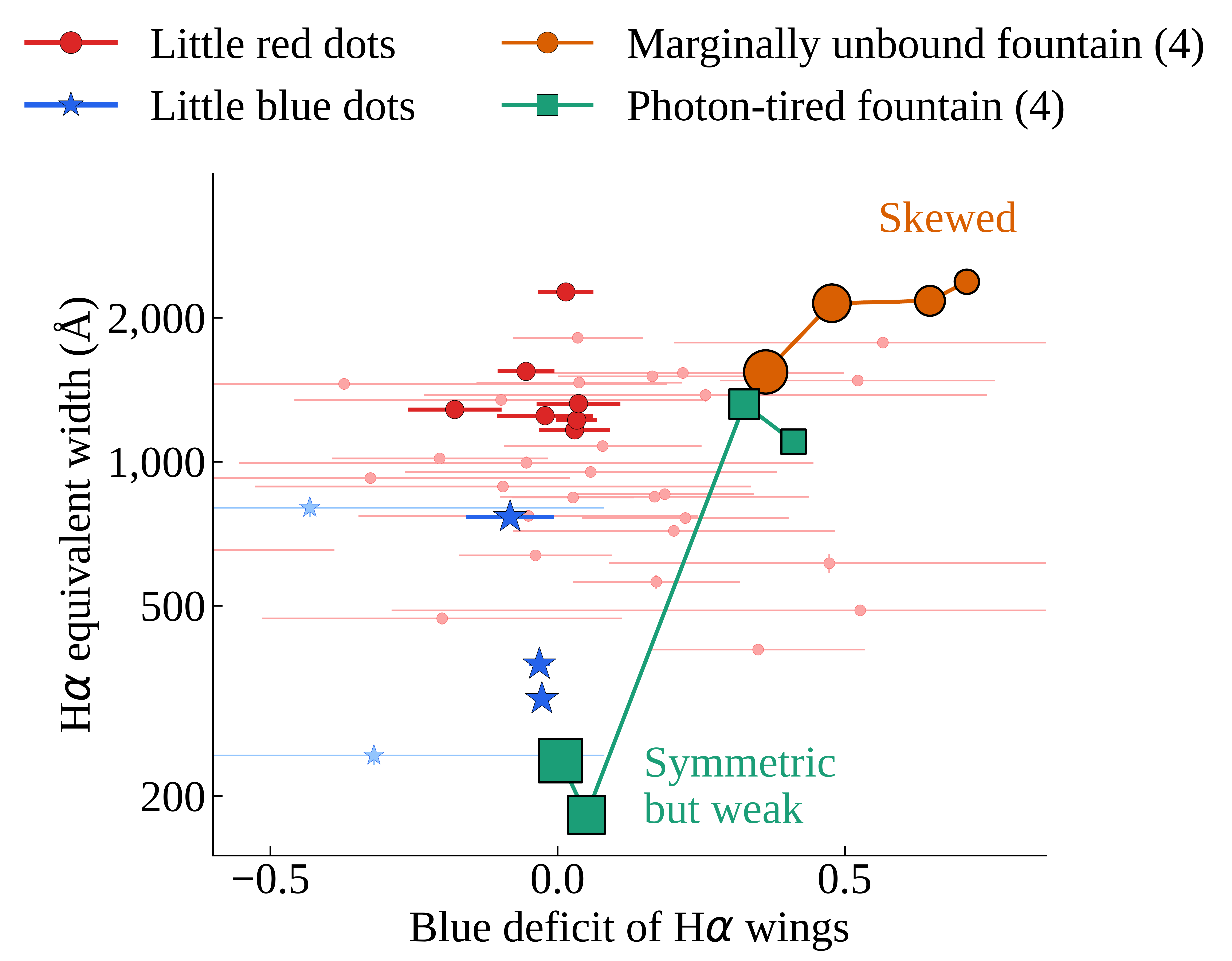}
\caption*{\textbf{Fig. 4 $|$ Published fountain spectra do not reproduce both the strength and symmetry of \Ha.} \Ha\ equivalent width versus blue-wing deficit, measured with identical windows in the observations and the eight published fountain spectra\cite{kaaz26}. Dark points denote sources with deficit uncertainties below 0.1.}\end{figure}
\begin{figure*}[t]\centering\includegraphics[width=\textwidth]{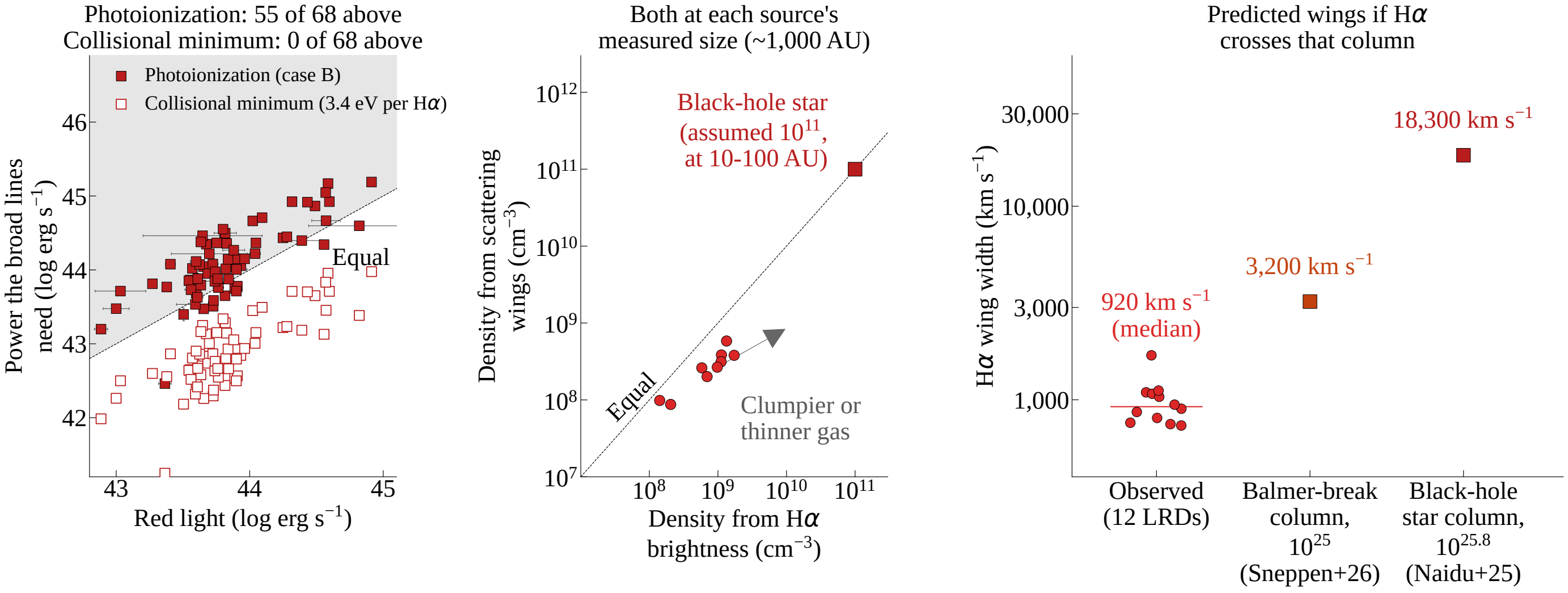}
\caption*{\textbf{Fig. 5 $|$ Black-hole-star and quasi-star envelopes fail on density and wing width.} Left: power required by the broad lines versus red-continuum luminosity, for case-B photoionization (filled) and for the collisional minimum of 3.4 eV per \Ha\ photon (open); the dashed line marks equality. Middle: gas density from the \Ha\ emission measure over the photospheric volume, and electron density from the scattering-wing optical depth over $0.3R$, both at each source's blackbody radius, compared with the black-hole-star model density\cite{naidu25}. Right: observed \Ha\ wing widths compared with those predicted from the column densities adopted in the Balmer-break\cite{sneppen26} and black-hole-star\cite{naidu25} fits.}\end{figure*}

Super-Eddington models must account for the energy released by rapid accretion. The proposed routes, namely winds, photon trapping and reprocessing in a massive envelope, make distinct spectral predictions.

A fast wind could escape outside our sightline. The fountain model\cite{kaaz26}, a preprint combining one-dimensional wind solutions with Monte Carlo radiative transfer, launches winds at $10^{3}$--$10^{4}$ \kms\ or more from a super-Eddington disc, much faster than the observed LRD absorbers. The fast flow escapes near the poles; a slow, marginally unbound wind covers most sightlines. Some gas escapes and some falls back, and the combined inflow and outflow can symmetrize the scattering wings. The outflow rate sets the spectrum: about 2.5 $M_\odot$\,yr$^{-1}$ produces an LBD, whereas 5--15 $M_\odot$\,yr$^{-1}$ produces an LRD with a partial or complete Balmer break.

We measured the eight simulated spectra published with the fountain model, none of which corresponds to an observed source, with the same \Ha\ wing-deficit and equivalent-width windows as the data (Fig.~4). Six have red-skewed wings, with blue deficits of 0.33--0.71. The two symmetric profiles have \Ha\ equivalent widths of only 182 and 237 \AA. The seven LRDs with the best-measured deficits have equivalent widths of 1{,}170--2{,}260 \AA, five to twelve times larger, and deficits consistent with zero. The two brightest LBDs have deficits of $-0.03\pm0.01$ and $-0.03\pm0.02$, at equivalent widths of 319 and 377 \AA. No published fountain spectrum reproduces both the observed strength and symmetry of \Ha.

Photon trapping does not remove the wind. Three-dimensional radiation-magnetohydrodynamic simulations of super-Eddington discs still radiate 3--5\% of the accreted rest-mass energy and expel about 30\% of the accreted mass\cite{jiang14,sadowski16}. These trapping flows predict the winds that the host-gas and nuclear diagnostics exclude.

Reprocessing in a massive envelope offers a third route. Quasi-stars embed a black hole in a radiation-supported envelope\cite{begelman08}; black-hole-star models use dense gas to produce the Balmer break and red continuum\citep{naidu25}. In both, a photosphere near 4{,}000--5{,}000 K emits too few photons to photoionize hydrogen. We modelled the red continuum as a blackbody in the 68 broad-line LRDs whose PRISM spectra constrain its temperature (Methods). Case-B photoionization requires more than the entire red-continuum luminosity in 55 of these 68, with a median ratio of 1.9 (Fig.~5), and photoionization from the $n=2$ level falls short of the photospheric Balmer-continuum photons by a median factor of 10 in 64 of 68. Collisional ionization in dense gas can power the lines with a median 12\% of the photospheric luminosity, but only at temperatures above about 6{,}000 K or densities near $10^{11}$ cm$^{-3}$, where models produce Balmer emission through a non-thermal $n=2$ population\cite{lrdcloudy26}. The fitted photospheres have a median temperature of 4{,}200 K, at which hydrogen in local thermodynamic equilibrium is less than 1\% ionized at the densities measured below.

Density and scattering provide two further tests. At the blackbody photospheric radii, about 1{,}000 AU, the \Ha\ emission measure implies a density of $1\times10^{9}$ cm$^{-3}$, whereas the electron column density inferred from the scattering wings gives $3\times10^{8}$ cm$^{-3}$ (Methods). The black-hole-star model requires $10^{11}$ cm$^{-3}$ at 10--100 AU\citep{naidu25}. The published Balmer-break\citep{sneppen26} and black-hole-star\citep{naidu25} fits adopt column densities of $10^{25}$ and $10^{25.8}$ cm$^{-2}$, respectively. The electron-scattering relation\cite{rusakov26} then predicts \Ha\ wing widths of 3{,}200 and 18{,}300 \kms, compared with the observed median of 920 \kms. None of the three routes by which a super-Eddington black hole could dispose of its excess energy matches the spectra.

\subsection*{Published outflows and jets}
Two further observations bear on these results: an outflow reported in one LRD, and jets as an alternative feedback channel. We test each in turn.

Earlier studies reported outflows in individual LRDs, most prominently GLIMPSED-329380 at $z=6.64$, whose broad \oiii\ wing has been interpreted as an outflow\cite{korber26}. We analysed its calibrated G395M spectrum separately from the main sample (Methods; Extended Data Fig.~1). Two narrow \oiii\ systems separated by $208\pm43$ \kms\ and $0.64\pm0.17$ kpc, plus a broad base, reproduce the profile to the noise level. We recover the published base at $-614\pm158$ \kms\ with FWHM 4{,}502 \kms. It lies $1.39\pm0.08$ kpc from the black hole traced by broad \Ha, shows no centroid shift across its wings and has no counterpart in [Ne\,\textsc{iii}] or [S\,\textsc{ii}]. The base is offset from the black hole by more than a kiloparsec and shows neither the velocity gradient nor the counterparts in other lines expected of a nuclear wind. Extended Data Table~1 gives our reanalysis of the other published LRD outflow claims at $z>4$ and identifies those for which the necessary data were unavailable.

Radio limits exclude powerful jets. Individual and stacked \textit{Chandra} observations place most LRDs well below the X-ray luminosities expected from their broad lines\cite{ananna24,yue24,maiolino25}, and radio surveys detect almost none of them\cite{mazzolari24,akins25,perger25,gloudemans25}. We converted the stacked 5 GHz limit of $2\times10^{39}$ erg s$^{-1}$ for broad-line AGN at $z\sim5$\cite{mazzolari24} to 1.4 GHz with a spectral index of 0.7, then applied the local relation between radio luminosity and mechanical jet power\cite{cavagnolo10}. The inferred jet power is below $2\times10^{43}$ erg s$^{-1}$, or 2\% of the median bolometric luminosity of our sample, as in local radio-quiet AGN. Weak compact jets can drive local outflows in radio-quiet AGN\cite{singha23}; our radio constraint concerns powerful jets. LRDs therefore drive neither superwinds nor powerful jets.

Our wind searches have three limits. First, winds broader than 1{,}500 \kms\ make up 3\% of the local population. Injected components with FWHM of 3{,}000 and 5{,}000 \kms\ are recovered in at most 3 of our 67 spectra; a separate search finds no such component, but this low recovery prevents their exclusion (Methods). Second, hot, neutral and molecular outflows are not traced by \oiii\ and require X-ray\cite{tombesi10}, sodium-absorption\cite{rupke05} or submillimetre\cite{cicone14} measurements. Third, the limits apply to sources bright enough for grating spectra. The broad-line and absorption diagnostics extend the test to the nuclear gas, which is bound in every LRD and LBD tested.

\section*{Discussion}

If LRDs and LBDs accreted like local AGN, a third to a half of them would drive \oiii\ winds. Below the Eddington limit, local AGN drive such winds in 31\% of cases, which predicts at least seven detections in our spectra; we find none (Fig.~1). Neither accretion regime explains this departure. Above the Eddington limit, the excess energy must leave as a wind, be trapped or be reprocessed in an envelope, and each of these routes conflicts with a measured property of the spectra. The data therefore point to radiatively efficient accretion below the Eddington limit, consistent with the dynamical black-hole mass measured in one LRD\cite{juodzbalis26}. Local AGN at these sub-Eddington rates drive winds; LRDs and LBDs at the same rates drive none. Their radiation fails to couple to the surrounding ionized gas as it does locally. Our data establish this missing coupling; its physical cause is the next question.

This result bears directly on galaxy formation. Feedback models link black-hole growth to the removal of gas that fuels both the nucleus and its host\cite{silk98,dimatteo05}. During the first 1.5 billion years, the LRDs and LBDs studied here drive none of the \oiii\ winds expected from nearby AGN: for winds narrower than 1{,}500 \kms\ with the local sub-Eddington distribution, the incidence is below 13\% at 95\% confidence. No wind-driven quenching is under way in these sources at the epoch we observe.

The missing winds also bear on how black holes and their hosts reach the local scaling relations. In the model of Silk and Rees\cite{silk98}, a black hole grows until its wind can expel the gas of its host, which sets the black-hole mass at a given stellar velocity dispersion. The black holes of LRDs lie above the local black-hole--stellar mass relation\cite{pacucci23}. None of our sources drives the wind that halts growth in this model, so wind-driven self-regulation is absent in them. Gas that a local AGN would expel stays available to the nucleus, and the black hole keeps accreting while its host is still small. Maiolino et al.\cite{maiolino25} described LRDs as feedback-free from their X-ray weakness. Our spectra test the winds directly, from the broad-line region to kiloparsec-scale ionized gas, and find none.

The wind-free phase we measure is common: LRDs outnumber ultraviolet-selected quasars of similar luminosity by one to two orders of magnitude\cite{matthee24,greene24}, so a large share of the black holes growing in the first 1.5 billion years grow without wind feedback. They enter later epochs already overmassive relative to their hosts, and the local relations must be set afterwards, as hosts build stellar mass and winds switch on. This sequence requires LRDs to evolve into the ordinary broad-line AGN that dominate at $z<4$, where LRDs become rare\cite{kocevski25}. Two measurements will test it. The offset of broad-line AGN above the local black-hole--stellar mass relation should shrink as ionized winds appear, and the wind incidence, measured with the same calibrated search, should rise from the LRD epoch towards cosmic noon. Applying this search to broad-line AGN at $z<4$ will show when the transition happens.

\section*{Methods}
Uncertainties are $1\sigma$ unless stated otherwise.

\subsection*{Sample}
Table~1 summarises the selection and the sample used for each analysis. We used version 4.4 of the DAWN JWST Archive, containing 80{,}367 NIRSpec spectra\cite{heintz25,valentino25}. The initial PRISM sample comprises 2{,}169 sources at $4<z<12$ with visually confirmed redshifts (grade 3) and photometric flux radii (\textit{flux\_radius} in the archive photometric catalogue\cite{dja_v7}) below 3 pixels, equivalent to 0.12$^{\prime\prime}$ on the 0.04$^{\prime\prime}$ archive-mosaic grid. We fitted the rest-frame ultraviolet (1{,}300--3{,}500 \AA) and optical (3{,}700--5{,}500 \AA) continua with power laws, $f_\lambda\propto\lambda^{\beta}$, after masking emission lines. Following \citet{kocevski25}, we defined a V-shaped continuum by $\beta_{\rm UV}<-0.37$ and $\beta_{\rm opt}>0$, selecting 242 sources. The remaining 1{,}927 compact sources form the LBD parent sample. They have blue or flat optical continua ($\beta_{\rm opt}\leq0$), red ultraviolet continua, or slopes that cannot be measured.

Published photometric selections\cite{kocevski25,greene24} include LRD candidates with PRISM spectra too faint for slope measurements or with grating spectra only. We therefore added 670 published candidates and matched them to the archive. Together with the V-shaped sources, these yield 334 LRD targets: 168 selected only by their V-shaped spectra, 92 only from the literature and 74 by both routes. We compare models with the Bayesian information criterion, ${\rm BIC}=\chi^{2}+k\ln n$, where $k$ is the number of free parameters and $n$ the number of pixels. A broad line is detected when adding a broad Gaussian to the narrow-line model lowers the BIC by more than 10, with FWHM of at least 1{,}000 \kms, a width resolved by the instrument, amplitude signal-to-noise ratio above 3 and no parameter at a bound; a BIC improvement of 6--10 counts as marginal. We apply this test to \Ha\ and \Hb\ separately. We detect broad Balmer lines in 99 LRDs, with 20 further marginal detections: 94 in \Ha, 17 in \Hb\ and 12 in both. The primary sample contains 95 of the 99 detections. We exclude two lensed counter-images of Abell 2744-QSO1 (LRDG-0089 and LRDG-0124), LRDG-0108, whose broad-line width is unconstrained (FWHM 8{,}500--23{,}200 \kms), and LRDG-0143, whose \Ha\ FWHM exceeds 30{,}000 \kms; a line this broad cannot be separated from the continuum within the fitting window, so its width and mass are unconstrained.

The LBD parent sample contains 47 broad-Balmer-line detections (46 in \Ha, 4 in \Hb\ and 3 in both) and 32 marginal detections. Twenty-seven of these 79 sources are published LRDs that fail our PRISM V-shape criterion, 19 with detected and 8 with marginal broad lines. We count them once, as LRDs, leaving 52 LBDs. Of these, 48 have $\beta_{\rm opt}\leq0$. The other four (LBDG-0089, LBDG-0106, LBDG-0263 and LBDG-0309) have PRISM spectra too faint in the rest-frame optical continuum for a slope measurement; three have blue ultraviolet slopes ($\beta_{\rm UV}=-2.0$ to $-3.4$). We retain them as LBDs because they are compact and absent from every published LRD list, and flag them in the source table. Two of them, LBDG-0106 and LBDG-0309, enter the wind search; excluding both lowers the expected detections from 7.2 and 7.5 to 6.9 and 7.2. LBDG-0106 is also the LBD absorber. Seven of the eight marginal cases have grating \oiii\ coverage and five enter the wind search as LRDs.

\subsection*{Spectra, noise and instrumental resolution}
We used the archive's one-dimensional msaexp extractions with their point-source path-loss correction. Each model component was convolved with the wavelength-dependent NIRSpec line-spread function and sampled at five sub-pixel positions. A free factor allowed the instrumental width to vary between 0.5 and 1.0 of its nominal value. We measured the noise in 10-pixel blocks of line-free continuum outside each fitting window and rescaled the pipeline errors by the ratio of measured to expected scatter. The median rescaling factor for grating spectra was 1.14. Fits minimised $\chi^{2}$ using bounded least squares from several initial conditions. Parameter uncertainties were scaled by the reduced $\chi^{2}$ when it exceeded unity. Models were compared with the BIC defined above. Unless stated otherwise, all line components were treated as emission.

\subsection*{\oiii\ decomposition and wind criterion}
Many sources were observed with more than one grating or programme. For the \oiii\ search we used one spectrum per source, the grating spectrum with the highest \oiii\ signal-to-noise ratio. The Balmer-line analyses below use the grating spectra that cover \Ha. We fitted rest-frame 4{,}420--5{,}450 \AA, masking 4{,}640--4{,}730 \AA. The base model comprised a linear continuum; the \oiii\ $\lambda\lambda4959,5007$ doublet with a fixed ratio of 1:2.98\cite{storey00}; narrow \Hb\ tied to the \oiii\ kinematics; a broad \Hb\ Gaussian with FWHM 800--8{,}000 \kms\ and centre within $\pm600$ \kms; and the optical Fe\,\textsc{ii} template\cite{boroson92}, broadened to a free FWHM of 900--5{,}000 \kms\ with its velocity tied to the broad Balmer line. The narrow \oiii\ component had a centre within $\pm250$ \kms\ of the line peak and FWHM 30--500 \kms. The wind model added a second \oiii\ doublet with FWHM 300--2{,}500 \kms\ and a shift of $-1{,}500$ to $+300$ \kms\ relative to the narrow core.

We tested whether the second component improves the fit with an $F$-test on the reduction in $\chi^2$, with 3 and $n-15$ degrees of freedom, converting the false-alarm probability to a Gaussian significance. A spectrum entered the usable sample if \oiii\ was detected above five standard deviations. A second component was significant if its fit improvement exceeded three standard deviations in this $F$-test, its amplitude exceeded three times its uncertainty and no parameter lay at a bound. A statistical wind candidate additionally required a blueshift from the narrow core greater than three times its velocity uncertainty. A second component therefore counts as a wind only if it is significant and blueshifted\cite{heckman81,singha22}. A redshifted component can trace inflow or rotation and is not counted, unless a blueshifted counterpart is also present, as an unresolved bipolar flow would produce. These cuts were unchanged between the primary search, injections and local comparison; no escape-speed or additional velocity cut was imposed on that statistical selection.

A statistically significant blueshift and an escaping superwind are distinct classifications. Following galaxy-formation models\cite{silk98,benson03}, we use superwind to denote gas expelled beyond the escape speed. We compare broad-line and absorber velocities with the escape speed at the adopted broad-line-region location, and host-gas \oiii\ kinematics with the host velocity dispersion. Following Maiolino et al.\cite{maiolino24}, we measured it for each source from the intrinsic width of the narrow \oiii\ core, $\sigma_\star=10^{0.15}\,{\rm FWHM}/2.355$, where the factor converts gas to stellar dispersion. The values span 38--125 \kms\ (16th--84th percentiles). The local $M_{\rm BH}$--$\sigma_\star$ relation\cite{kormendy13} gives values a median 1.6 times higher; we use the direct measurement. These comparisons inform the physical interpretation; they do not alter the primary detection count or recovery calibration.

Among the six significant primary components, only NEW-0011 is blueshifted, at $-47\pm19$ \kms, below the three-standard-deviation threshold. Four LRDs have redshifts of 26--88 \kms; the remaining LRD component lies at $+240\pm130$ \kms\ with FWHM 1{,}290 \kms.

The two supplementary G395H refits must be distinguished from their primary measurements. The primary G395M fit to LRDZ J141910.88+524719.8 gives $-58\pm45$ \kms\ and amplitude signal-to-noise ratio 2.2, failing the amplitude cut. The primary G235M fit to LRDZ J123654.99+620846.2 gives $+26\pm28$ \kms, a fit improvement of 7.0 standard deviations in the $F$-test and amplitude signal-to-noise ratio 3.4: it is a significant redshifted component in the six-source group. This significance measures how much the second component improves the fit, which depends on its flux and width, not its shift: this component has FWHM 770 \kms, nearly three times the 270 \kms\ core, at almost the same velocity. The supplementary G395H values are $-72\pm12$ \kms\ with FWHM $466\pm25$ \kms\ and, with Fe\,\textsc{ii} included, $-38\pm9$ \kms\ with FWHM 320 \kms, respectively. Their blueshifts have significances of 6.0 and 4.2 standard deviations. Neither supplementary fit replaced a primary fit or entered the recovery calculation. They reveal slow blueshifted gas, while the reported zero count refers specifically to the primary search.

\subsection*{Injection and recovery}
We injected a synthetic second \oiii\ doublet into each of the 67 usable spectra over a grid of four FWHM values (400, 600, 1{,}000 and 1{,}500 \kms), five velocity shifts ($-200$, $-400$, $-600$, $-900$ and $-1{,}200$ \kms) and six fractions of the total \oiii\ flux (3, 5, 10, 20, 30 and 40\%). This gives 120 injections per spectrum and 8{,}040 in total. Each injected spectrum was refitted from scratch with the base and wind models. An injection was recovered only if it passed the same wind criterion as the real data. These tests define a recovery function for each source in FWHM, shift and flux fraction.

\subsection*{Local comparison samples}
The local \oiii\ comparison draws on three samples. We fitted 208 type 1 AGN from BASS DR1\cite{koss17} and 23 SDSS DR7 AGN with continuum-based $L_{\rm bol}\geq L_{\rm Edd}$\cite{liu19} with the same model and wind criterion used for the high-redshift sample. For the 9{,}455 type 1 AGN in ALPAKA\cite{alpaka}, also drawn from SDSS DR7, we applied that criterion to the published two-component \oiii\ decompositions. One of the 23 SDSS AGN is also in ALPAKA. All SDSS spectra used here are DR7 legacy spectra, retrieved from the DR17 archive. Black-hole masses were estimated from the broad Balmer lines\cite{greene05}, using ALPAKA broad \Ha\ widths from \citet{liu19}. Eddington ratios were calculated with $L_{\rm bol}=3500\,L_{\rm [O\,III]}$\cite{heckman04}, using only the narrow component so that wind emission did not affect the classification. In these type 1 AGN the narrow \oiii\ is dominated by the nucleus; the continuum-based check below, which does not use \oiii, gives the same result. Four sources, three in ALPAKA and one in SDSS, lacked a black-hole mass or narrow \oiii\ luminosity and were excluded. The remaining 9{,}682 sources (208 BASS, 22 SDSS and 9{,}452 ALPAKA) comprise 412 super-Eddington AGN, 203 with winds, and 9{,}270 sub-Eddington AGN, 2{,}908 with winds. The Eddington division is made on this \oiii\ scale for all three samples, including BASS. Every local wind has $\sigma_{\rm wind}/\sigma_\star>1$, with $\sigma_\star$ inferred from the $M_{\rm BH}$--$\sigma_\star$ relation, which is calibrated on local galaxies\cite{kormendy13}.

We passed each local wind through each high-redshift source's recovery function, interpolating in FWHM, shift and flux fraction. Winds broader than 1{,}500 \kms\ or slower than 200 \kms\ were counted as undetectable, making the predicted detection counts lower limits. The broad-width exclusion affects 3\% of local winds; none exceeds 2{,}500 \kms. If $p_i$ is the mean recovery probability for source $i$ and $f$ the local wind fraction, the expected number of detections is $\sum_i f p_i$ and the probability of zero detections is $\prod_i(1-f p_i)$. As a check on the \oiii-based division, we also used Eddington ratios from the host-subtracted 5{,}100 \AA\ continuum\cite{liu19}, available for 8{,}133 ALPAKA AGN. This selects 442 super-Eddington AGN, 197 with winds, and predicts 11.2 detections, with $P_0=4\times10^{-6}$.

The local samples differ from ours in black-hole mass and luminosity, and local wind incidence depends on both. We therefore reweighted each local sample to the distribution of our sources in black-hole mass and bolometric luminosity, in bins of 0.5 dex (Extended Data Fig.~2). For sub-Eddington AGN, matching in both quantities covers 65 of the 67 sources and gives a wind fraction of 34\%, 8.0 expected detections and $P_0=2\times10^{-4}$. Local super-Eddington AGN are more luminous than our sources at the same mass by definition, so we matched them in mass alone: the wind fraction is 49\%, with 5.8 expected detections and $P_0=2\times10^{-3}$. Matching the sub-Eddington sample in mass alone gives 6.9 detections and $P_0=6\times10^{-4}$.

The bolometric luminosities of our sources come from broad \Ha, through $L_{\rm bol}=9.26\,L_{5100}$ with $L_{5100}$ inferred from the broad-line luminosity\cite{greene05}. We therefore repeated the comparison with the local bolometric luminosities derived the same way, from broad \Ha\ for ALPAKA and broad \Hb\ for BASS and SDSS\cite{greene05}. On this scale the local Eddington ratios are a median 0.19 dex lower than on the \oiii\ scale, and 8 of the 9{,}682 local AGN, and 2 of our 67 sources, exceed the Eddington limit (Extended Data Fig.~2c). Reweighted to the Eddington-ratio distribution of our sources, local AGN have a wind fraction of 37\%, which predicts 8.2 detections with $P_0=1\times10^{-4}$; matching also in black-hole mass gives 39\%, 9.1 detections and $P_0=4\times10^{-5}$. The 783 local AGN with $\lambda_{\rm Edd}\geq0.32$ on this scale drive winds in 46\% of cases and predict 9.2 detections. The wind deficit therefore does not depend on the luminosity tracer.

\subsection*{Search for very broad \oiii\ components}
We extended the search beyond the injection grid by allowing a second \oiii\ component with FWHM 1{,}500--9{,}000 \kms. This yielded no additional wind candidate among the 67 sources. We injected components with FWHM 3{,}000 and 5{,}000 \kms, a shift of $-600$ \kms\ and 10\% or 20\% of the \oiii\ flux into each spectrum. None was recovered at 10\%. At 20\%, the 3{,}000 \kms\ component was recovered in 3 of the 67 spectra and the 5{,}000 \kms\ component in 1; among the 17 spectra with \oiii\ signal-to-noise ratio above 50, the rates are 2 of 17 and 1 of 17. Our incidence limits therefore apply to winds narrower than 1{,}500 \kms.

As supplementary checks, we retrieved two G395H/F290LP spectra of LRDZ J141910.88+524719.8 from JWST programme 8410 (OCEANS) in the Mikulski Archive for Space Telescopes. Both require a second component, with fit improvements of 9.0 and 7.6 standard deviations. Their shifts are $-71\pm16$ and $-74\pm19$ \kms\ and their FWHM values are $509\pm37$ and $431\pm33$ \kms, respectively. We applied the same fitting procedure to the other LRDs at $z>4$ with published outflow or absorber claims and available public spectra (Extended Data Table~1). These source-specific checks are separate from the fixed primary-spectrum search and its 120 injections per source. The separate analysis of GLIMPSED-329380 is described below.

\subsection*{Reanalysis of GLIMPSED-329380}
Korber et al.\cite{korber26} fitted a broad blueshifted \oiii\ component in GLIMPSED-329380 at $z=6.64$ and interpreted it as a fast AGN-driven outflow, the strongest such claim in an LRD; Extended Data Fig.~1 shows our decomposition. We retrieved the pipeline one-dimensional G395M/F290LP spectrum of GLIMPSED-329380 (JWST programme 9223) from the Mikulski Archive for Space Telescopes and adopted the narrow-line redshift $z=6.63784$ from \citet{korber26}. Korber et al. extracted the host and the AGN separately and report the outflow in the host extraction. Separating a host from an unresolved AGN requires an accurate model of the AGN emission\cite{husemann13,singha22}; LRDs have no such model, and an imperfect separation transfers broad AGN emission into the host spectrum. The broadest component reported in their host extraction, with FWHM $\approx5{,}350$ \kms, is comparable to the broad base of our integrated fit (FWHM $4{,}502\pm492$ \kms). We therefore fitted the integrated spectrum, which contains all the emission and does not depend on that separation. The scatter in line-free regions at rest-frame 4{,}790--4{,}840, 5{,}060--5{,}150 and 5{,}190--5{,}250 \AA\ is 1.32 times the pipeline errors; we rescaled the errors accordingly. We fitted 4{,}780--5{,}250 \AA\ with a linear continuum and Gaussian line components convolved with the G395M line-spread function. Within each component, the \oiii\ doublet and \Hb\ share a centre and FWHM, with free amplitudes and a fixed \oiii\ doublet ratio of 1:2.98.

Double-peaked narrow lines are not unique to outflows; nearby compact galaxies also show profiles consistent with distinct nuclei\cite{lin23}. The starting model contains a narrow component with centre between $-300$ and $+400$ \kms\ and FWHM 50--600 \kms; a second component with FWHM 300--1{,}500 \kms; He\,\textsc{i} $\lambda4922$ fixed at 0.1 times the measured He\,\textsc{i} $\lambda5876$ flux; and a broad \Hb\ Gaussian. We tested every combination of four additions: a second narrow component with the same bounds as the first; a broad base with FWHM 1{,}500--6{,}000 \kms; Fe\,\textsc{ii} $\lambda\lambda4924,5018,5169$ tied kinematically to broad \Hb; and a blueshifted \oiii\ component at $-2{,}500$ to $-800$ \kms.

The lowest BIC is obtained with two narrow components, the second component and the broad base ($\chi^2_\nu=1.03$, 21 parameters). The narrow components have velocities of $+123\pm12$ and $-85\pm42$ \kms\ and FWHM values of $134\pm28$ and $102\pm97$ \kms. Removing the second narrow component raises the BIC by 178. Relative to the brighter narrow component, the second component lies at $+4\pm6$ \kms\ with FWHM $633\pm18$ \kms, and the base lies at $-614\pm158$ \kms\ with FWHM $4{,}502\pm492$ \kms\ and signal-to-noise ratio 11. Adding Fe\,\textsc{ii} or the extra blueshifted component raises the BIC by 13.5 and 15.9, respectively. Parameter uncertainties were taken from the covariance matrix.

Broad-line emission provides a spatial reference for spectroastrometric measurements of ionized gas in type 1 AGN\cite{singha22,winkel23}. To locate the components spatially, we initially fixed their centres and widths to the spectral-fit values, fitted their amplitudes and the continuum independently in each row of the two-dimensional spectrum, and fitted a Gaussian to each resulting spatial amplitude profile. We propagated decomposition uncertainties by repeating this procedure 300 times, drawing centres and widths from their uncertainties. The two narrow components are separated by $1.17\pm0.31$ pixels, or $0.64\pm0.17$ kpc in the image plane at 0.1 arcsec per pixel. The base lies $0.32\pm0.37$ pixels from the brighter narrow component.

We applied the same spatial analysis to \Ha, [N\,\textsc{ii}] and [S\,\textsc{ii}]. The broad \Ha\ Gaussian has FWHM $3{,}389\pm532$ \kms\ and a velocity of $-180\pm226$ \kms\ relative to the brighter narrow component. Its spatial offset from that component is $2.52\pm0.14$ pixels, or $1.39\pm0.08$ kpc. The brighter narrow component has consistent positions near \Hb\ and \Ha, differing by $0.07\pm0.29$ pixels, so a tilt in the spectral trace does not account for the offset.

To test for a velocity gradient across the base, we subtracted the fitted continuum, narrow components and second component from each spatial row. We then measured the spatial centroid of the remaining \oiii\ $\lambda5007$ flux in velocity bins, estimating uncertainties from 400 resamplings of the wavelength pixels. The bins at $-2{,}400$ to $-1{,}300$, $-1{,}300$ to $-500$ and $+500$ to $+1{,}300$ \kms\ have flux signal-to-noise ratios of 3.7, 3.1 and 3.5, and centroids of $14.79\pm0.20$, $14.79\pm0.34$ and $14.44\pm0.25$ pixels, respectively. The blue-to-red offsets are $0.35\pm0.32$ and $0.35\pm0.42$ pixels.

Finally, we fitted \Ha\ and [N\,\textsc{ii}] with narrow and second components plus a broad \Ha\ Gaussian. We fitted [Ne\,\textsc{iii}] $\lambda3869$, masking He\,\textsc{i} $\lambda3889$ and H8, and [S\,\textsc{ii}] $\lambda\lambda6718,6733$ with narrow and second components. Neither [Ne\,\textsc{iii}] nor [S\,\textsc{ii}] requires the second component ($\Delta{\rm BIC}<0$).

\subsection*{Broad Balmer lines}
We first decomposed \Ha\ into continuum, narrow lines and broad emission. After subtracting the continuum and narrow components, we fitted the broad residual within $\pm3.5$ FWHM of its peak. The base profile is a Gaussian convolved with a two-sided exponential\cite{rusakov26}, evaluated on a 25 \kms\ grid. Its Gaussian dispersion spans 30--6{,}000 \kms, its exponential scale spans 50--15{,}000 \kms, and its centre is free within $\pm3{,}000$ \kms\ of the peak. The outflow test adds a Gaussian with FWHM 500--5{,}000 \kms\ and a shift of $-3{,}000$ to $-100$ \kms\ from the base centre. The control uses the same Gaussian with a shift of $+100$ to $+3{,}000$ \kms. We rescaled $\Delta\chi^{2}$ by the reduced $\chi^{2}$ of the base fit and required $\Delta{\rm BIC}>10$, amplitude and shift greater than three times their respective uncertainties, and no parameter within 20 \kms\ of a bound. In the two-component version, the base consists of two independent scattering profiles before the extra Gaussian is added. Local broad \Ha\ profiles are better described by two Gaussians than by one scattering profile ($\Delta{\rm BIC}>10$ in 88\% of the 601 SDSS AGN), so we repeated the test with a two-Gaussian base, the standard decomposition for local AGN. The blueshifted component is then required in 114 of 601 local AGN (19\%) and 6 of 47 LRDs (13\%; Fisher $p=0.34$), and a redshifted one in 95 and 3, respectively. The base profile therefore does not set the result.

The samples comprise 47 LRDs with broad \Ha\ signal-to-noise ratio above 5, seven LBDs and 601 SDSS broad-line AGN, all fitted with the same code. The \oiii\ comparison sample cannot serve here, because this test needs our own fits to high signal-to-noise broad \Ha\ profiles, whereas ALPAKA supplies only published \oiii\ decompositions. We therefore drew the 601 AGN from the same SDSS DR7 parent sample\cite{liu19}, keeping those with broad \Ha\ signal-to-noise ratio above 40. \Hb\ enters only through the catalogue values of $R_{\rm Fe\,II}$ and FWHM(\Hb) used to place each source on the quasar main sequence; the test itself uses \Ha\ alone, as for the LRDs. The SDSS sources sample the two ends of the main sequence: 325 have strong Fe\,\textsc{ii} ($R_{\rm Fe\,II}\geq1$) and FWHM(\Hb) $<4{,}000$ \kms, with median $\lambda_{\rm Edd}=0.18$; 276 have weak Fe\,\textsc{ii} ($R_{\rm Fe\,II}<0.5$) and FWHM(\Hb) $>4{,}000$ \kms, with median $\lambda_{\rm Edd}=0.06$. We estimated these Eddington ratios from our broad \Ha\ fits, deriving the mass from the line width and luminosity\cite{greene05} and adopting $L_{\rm bol}=9.26\,L_{5100}$, with $L_{5100}$ inferred from the broad \Ha\ luminosity\cite{greene05}; this is the same recipe used for the LRDs. Local wind incidence and blue-component incidence both rise with black-hole mass, so we reweighted the 538 SDSS AGN with a catalogue mass to the LRD mass distribution in 0.5 dex bins. The blue fraction becomes 22\% (two-Gaussian base: 19\%), compared with 17\% (13\%) in LRDs, with Fisher $p=0.57$ (0.42) for the effective local sample size. The broad \Ha\ signal-to-noise threshold is above 5 for LRDs and LBDs and above 40 for SDSS, whose larger sample allows us to retain only the best-measured profiles. The kinematic test compares each detected component's speed with its source's escape speed; the ratio itself is independent of this selection threshold.

For a virial broad-line region, $GM/R=f\,{\rm FWHM}^{2}$ and $v_{\rm esc}=\sqrt{2f}\,{\rm FWHM}$. This is the escape speed from the black hole at the broad-line radius, where the black hole dominates the potential; the host only adds to the binding, so the threshold is a lower limit. We adopt $f=0.5$, giving $v_{\rm esc}={\rm FWHM}$. This is lower than both the isotropic value $f=0.75$ adopted in the mass calibration we use\cite{greene05} and the value calibrated against the local $M_{\rm BH}$--$\sigma_\star$ relation for FWHM-based masses, $f\approx1.1$\cite{woo15}. A lower $f$ gives a lower escape speed, so $f=0.5$ is the conservative choice for the confinement test. Since $v_{\rm esc}\propto f^{1/2}$, the result is insensitive to $f$: for $f=0.75$, 1 and 1.5, no LRD component and at most two local components exceed the escape speed. Only for $f=0.25$, half our adopted value, does one LRD component reach it ($1.02\,v_{\rm esc}$), along with 7 of 148 local components. We define the broad-line centroid as the base-profile centre relative to the narrow lines. Its uncertainty combines 200 realisations of the rescaled noise with the fitted centre error and is 15--23 \kms. LRDZ J141910.88+524719.8 (LRDG-0054) is plotted at its fitted component speed, 0.72\,\vesc.

\subsection*{Balmer absorbers}
We searched every grating \Ha\ fit for absorption by multiplying the continuum and broad emission by a Gaussian absorption trough and keeping it when the BIC improved by more than 10. This gave 17 LRD candidates and one LBD candidate. We then fitted \Ha\ and \Hb\ jointly. The absorption multiplies the continuum and broad emission by $1-C_{f}[1-\exp(-\tau_0\phi_v)]$, leaving the narrow lines unabsorbed: they arise on kiloparsec scales, outside the dense gas that produces $n=2$ absorption. Here $\tau_0$ is the line-centre optical depth, $\phi_v$ is a Gaussian velocity profile with unit peak convolved with the line-spread function, and $C_f$ is the fraction of emission covered by the absorber. The two lines share absorber velocity, width and covering fraction. Atomic physics fixes the optical-depth ratio, $\tau({\rm H}\alpha)/\tau({\rm H}\beta)=f_{{\rm H}\alpha}\lambda_{{\rm H}\alpha}/(f_{{\rm H}\beta}\lambda_{{\rm H}\beta})=7.25$.

Of 17 candidate LRD absorbers, 14 survived the refit and 12 have \Hb\ coverage. Velocities are measured relative to the narrow-line redshift. The largest blueshift, $296\pm19$ \kms\ in LRDZ J141910.88+524719.8 (LRDG-0054), corresponds to $0.08\,v_{\rm esc}$. The two absorbers without \Hb\ coverage, LRDG-0052 and NEW-0011 (LRDG-0581), lie at $-81$ and $+26$ \kms, respectively. The LBD absorber, LBDG-0106, lies at 104 \kms.

The hydrogen column density in the $n=2$ level is $N_2=\tau_0 b/(1.497\times10^{-15} f_{{\rm H}\alpha}\lambda_{{\rm H}\alpha})$ cm$^{-2}$, where $b={\rm FWHM}/1.665$ is the Doppler parameter in \kms, $f_{{\rm H}\alpha}=0.641$ and $\lambda_{{\rm H}\alpha}=6{,}563$ \AA. We propagate the covariance between optical depth and width. Absorbers at the 50 \kms\ width floor are unresolved: their velocities are measured, but their column densities are unconstrained. The local comparison uses first-epoch measurements of six nearby AGN with Balmer absorption\cite{localabs}. For each fountain model, we divided the wind speed at every radius of the published solution by the local escape speed at that radius and averaged the ratio, weighting by the $n=2$ column.

\subsection*{Fountain spectra}
The fountain model\cite{kaaz26} combines one-dimensional, spherically symmetric wind solutions with Monte Carlo radiative transfer. It places a super-Eddington disc, viewed off-axis, inside a slow, marginally unbound wind, part of which falls back. We used its eight published simulated spectra; none corresponds to an observed source. They comprise marginally unbound and photon-tired winds with outflow rates of 2.5, 5, 10 and 15 $M_\odot$\,yr$^{-1}$. All models contain a $10^{6}\,M_\odot$ black hole, the only mass with published spectra; for the 15 $M_\odot$\,yr$^{-1}$ runs, we used the truncated outputs. Each model was smoothed to the resolution of each PRISM spectrum and sampled on its pixels. After subtracting continuum and narrow lines, we measured the blue deficit as the fractional difference between the mean flux at $-3{,}200$ to $-2{,}000$ \kms\ and the mirrored red interval. The \Ha\ equivalent width integrates the line within $\pm6{,}000$ \kms\ above a linear continuum fitted at 9{,}000--14{,}000 \kms\ on both sides. Models and observations were measured identically. For LBDs, the deficit was measured in the same velocity interval in the grating spectrum and the equivalent width in the matched PRISM spectrum. We excluded LBDG-0290 because its fit failed and LBDG-0285 because its deficit uncertainty is 8.5.

\subsection*{Ionizing power, density and wing width}
The quasi-star and black-hole-star interpretations identify the red continuum with a photosphere. We therefore fitted each PRISM spectrum above rest-frame 1{,}340 \AA, masking emission lines, with a blackbody for the red continuum and a power law for the blue light. The blackbody luminosity $L_{\rm BB}$ and temperature $T$ determine the radius through $L_{\rm BB}=4\pi R^{2}\sigma_{\rm SB}T^{4}$. We retained 68 broad-line LRDs from the primary sample. Each has PRISM coverage to rest-frame 6{,}900 \AA, a blackbody temperature between 1{,}600 and 19{,}000 K with uncertainty below 1{,}500 K, and a blackbody contribution above 30\% of the continuum at 5{,}500 \AA.

We inferred the required ionizing power from the broad-line luminosity. Under case-B recombination at $T=10^{4}$ K and $n_e=10^{4}$ cm$^{-3}$, each absorbed hydrogen-ionizing photon produces 0.45 \Ha\ photons\cite{osterbrock06}. Dividing the \Ha\ luminosity by $0.45\times1.889$ eV therefore gives the required absorbed ionizing-photon rate. For $L_\nu\propto\nu^{-1}$ between 13.6 eV and the He\,\textsc{ii} edge at 54.4 eV, the mean energy of an ionizing photon is $\ln4/(13.6^{-1}-54.4^{-1})=25$ eV. We multiplied the photon rate by this energy to obtain the ionizing power and compared it with $L_{\rm BB}$. Harder spectra raise the mean photon energy and the required power, so this choice is conservative. Where only \Hb\ is measured, we used $L_{{\rm H}\alpha}=3.1\,L_{{\rm H}\beta}$, the standard Balmer decrement of AGN\cite{osterbrock06}, without a dust correction. Case B assumes photoionization from the ground state. We therefore also computed two lower bounds. For photoionization from the $n=2$ level, each \Ha\ photon requires at least one photon above 3.4 eV; integrating each fitted blackbody gives a median of 1.4\% of its luminosity above this energy, and the required photon rate exceeds the available rate by a median factor of 10, in 64 of 68 LRDs. For collisional ionization from $n=2$, the minimum energy cost is 3.4 eV per \Ha\ photon, which requires a median 12\% of $L_{\rm BB}$ and exceeds it in no source. The Saha equation gives hydrogen ionized fractions of 0.5\% at the median blackbody temperature of 4{,}200 K and $n_{\rm H}=10^{9}$ cm$^{-3}$, 12\% at 5{,}000 K and 84\% at 6{,}000 K.

We estimated gas densities independently from the \Ha\ luminosity and the electron-scattering wings. For recombination emission, $L_{{\rm H}\alpha}=\epsilon\int n^{2}\,dV$, with $\epsilon=3.56\times10^{-25}$ erg cm$^{3}$ s$^{-1}$ at $10^{4}$ K. The emission measure is therefore $EM=L_{{\rm H}\alpha}/\epsilon$, in cm$^{-3}$. Gas filling the photospheric volume $V=4\pi R^{3}/3$ has density $n=(EM/V)^{1/2}$, in cm$^{-3}$.

The scattering estimate uses the 12 LRDs with broad \Ha\ signal-to-noise ratio above 40. The wing scale gives the scattering optical depth through $W=428\,\tau+370$ \kms\ at $10^{4}$ K\cite{rusakov26}. We converted this to an electron column density, $N_e=\tau/\sigma_{\rm T}$, and obtained a density by distributing the column density over $0.3R$. Distributing it over $0.1R$ or $R$ gives $9\times10^{8}$ or $9\times10^{7}$ cm$^{-3}$, still more than two orders of magnitude below the black-hole-star density. Applying the same width--optical-depth relation to the column densities in the published Balmer-break\cite{sneppen26} and black-hole-star\cite{naidu25} fits gives their predicted wing widths.

\subsection*{Use of AI tools}
The original idea for this work was developed by the authors. We used Claude Opus 5.5 (Anthropic) between 1 and 25 November 2026 to polish the text. The authors reviewed all edits and take full responsibility for the content of the manuscript.

\subsection*{Data availability}
The JWST spectra are public in the DAWN JWST Archive and the SDSS spectra in SDSS DR17. Source-by-source measurements, the injection--recovery results and a list of every source identifier used here (LRDG, LBDG, NEW and LRDZ names) with its coordinates, redshift and archive spectrum are available at Zenodo (\url{https://doi.org/10.5281/zenodo.24533555}).

\subsection*{Code availability}
The fitting, injection--recovery and figure scripts are available at Zenodo (\url{https://doi.org/10.5281/zenodo.24533555}).

\section*{Acknowledgements}
This work is based on observations made with the NASA/ESA/CSA James Webb Space Telescope. The data were obtained from the Mikulski Archive for Space Telescopes at the Space Telescope Science Institute, which is operated by the Association of Universities for Research in Astronomy, Inc., under NASA contract NAS 5-03127 for JWST. Some of the data products presented here were retrieved from the DAWN JWST Archive, which is funded by the Danish National Research Foundation under grant DNRF140. Funding for the Sloan Digital Sky Survey has been provided by the Alfred P. Sloan Foundation, the Participating Institutions, the National Science Foundation and the U.S. Department of Energy Office of Science. We thank the BAT AGN Spectroscopic Survey team for making their spectra public. M.S. acknowledges support from NASA under award number 80GSFC21M0002. Y.N. acknowledges funding from a Flatiron Research Fellowship. The Flatiron Institute is a division of the Simons Foundation. W.L.T. acknowledges support from NSF grant AST-2407089 and the Penn State Eberly Endowment.

\section*{Author contributions}
M.S. conceived the project, performed the analysis and wrote the manuscript. Y.N. contributed to the physical interpretation and the comparison with super-Eddington models. W.L.T. contributed to the interpretation and edited the manuscript. All authors discussed the results and commented on the manuscript.

\section*{Competing interests}
The authors declare no competing interests.

\section*{Additional information}
Correspondence and requests for materials should be addressed to Mainak Singha.

\renewcommand{\refname}{Methods references}
\makeatletter\let\oldthebib\thebibliography\renewcommand{\thebibliography}[1]{\oldthebib{#1}\setcounter{NAT@ctr}{45}}\makeatother

\clearpage
\begin{figure*}[t]\centering\includegraphics[width=\textwidth]{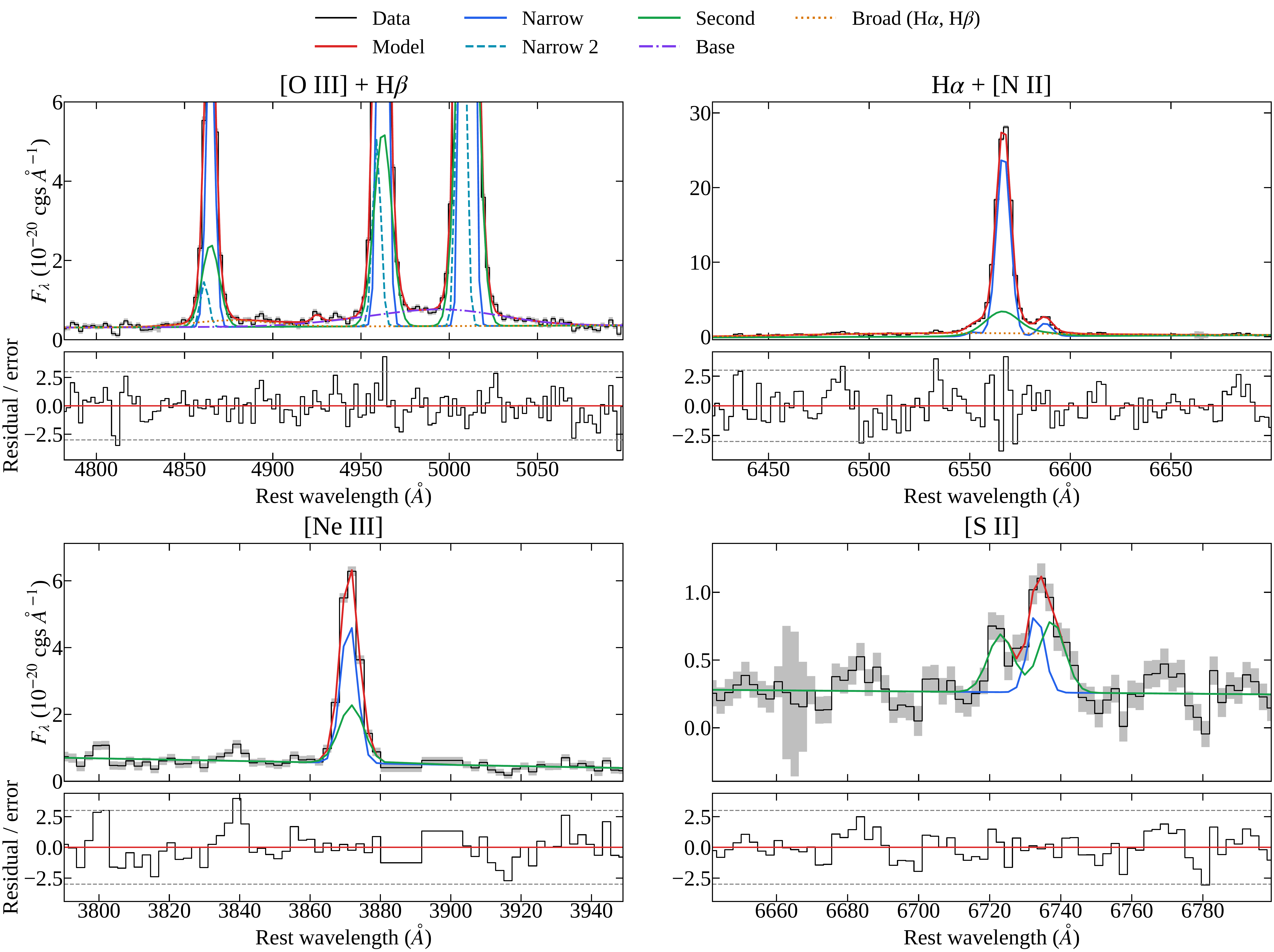}
\caption*{\textbf{Extended Data Fig. 1 $|$ Spectral decomposition of GLIMPSED-329380.} Each panel shows the public G395M spectrum (black), uncertainties (grey), fitted model (red), its individual components and residuals divided by the uncertainties. Upper left, \oiii\ and \Hb\ fitted with two narrow components, a second component and a broad base; the flux axis is truncated. The remaining panels use narrow and second components. Upper right, \Ha\ and [N\,\textsc{ii}], including broad \Ha. Lower left, [Ne\,\textsc{iii}] $\lambda3869$, with He\,\textsc{i} $\lambda3889$ and H8 masked. Lower right, [S\,\textsc{ii}] $\lambda\lambda6718,6733$.}\end{figure*}

\begin{figure*}[t]\centering\includegraphics[width=\textwidth]{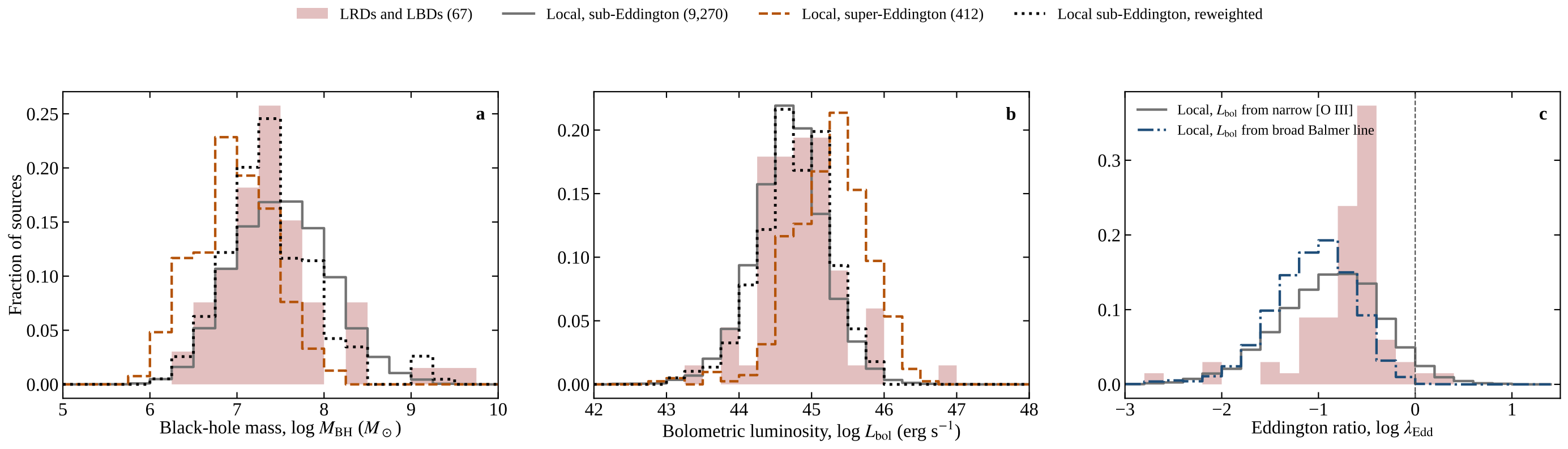}
\caption*{\textbf{Extended Data Fig. 2 $|$ Black-hole masses, luminosities and Eddington ratios of our sources and the local comparison.} \textbf{a}, Black-hole mass for the 67 LRDs and LBDs in the wind search (shaded), local sub-Eddington AGN (solid), local super-Eddington AGN (dashed) and the sub-Eddington AGN reweighted to our sources in mass and luminosity (dotted). \textbf{b}, The same for bolometric luminosity. \textbf{c}, Eddington ratio of our sources and of all 9{,}682 local AGN, with local bolometric luminosities from narrow \oiii\ (solid) or from the broad Balmer line (dash-dotted), the tracer used for our sources. The dashed vertical line marks the Eddington limit.}\end{figure*}

\begin{table*}[t]\centering\scriptsize\setlength{\tabcolsep}{3pt}
\caption*{\textbf{Extended Data Table 1 $|$ Reanalysis of published LRD outflow claims at $z>4$.} All velocities are in \kms. The \oiii\ fits cover rest-frame 4{,}900--5{,}070 \AA\ and include narrow and second \oiii\ components, free-width Fe\,\textsc{ii} $\lambda\lambda4924,5018$ Gaussians and He\,\textsc{i} $\lambda5016$, with the wind criterion defined in Methods. Our fits differ from the published ones by including Fe\,\textsc{ii} and by applying the same wind criterion to every source. The separate GLIMPSED-329380 analysis is described in Methods. \oiii\ velocities are relative to the narrow core; absorber entries give the \Ha\ absorption centroid. Our absorber velocities for RUBIES-182791, RUBIES-42046, SPURS-GN-2004 and OCEANS 100424 differ from the published values. Cases not reanalysed are identified in the final column.}
\begin{tabular}{@{}>{\raggedright\arraybackslash}p{3.0cm}l>{\raggedright\arraybackslash}p{3.6cm}>{\raggedright\arraybackslash}p{4.9cm}>{\raggedright\arraybackslash}p{2.6cm}@{}}\toprule
Object (other names) & $z$ & Published claim & This work & Outcome \\ \midrule
GLIMPSED-329380 & 6.64 & Fast [O III] outflow\cite{korber26} & Base at $-614\pm158$, FWHM 4{,}502 (Methods); no centroid shift across its wings & Recovered; base 1.4 kpc from the black hole \\
JADES-GN-2389 & 7.04 & [O III] outflow, $v_{\rm out}=758$\cite{cooper25} & [O III] $-156\pm83$, FWHM 635 & Recovered; blueshift $1.9\sigma$ \\
JADES-GS-159717 & 5.08 & [O III] outflow at $-20$\cite{deugenio25} & [O III] $-25\pm17$, at width bound; absorber $-16\pm7$ & Recovered; not blueshifted \\
COLA1 & 6.59 & Redshifted [O III], $+53$\cite{torralba26} & [O III] $+40\pm28$ (IFU); MSA data exclusive until 2027 & Recovered; redshifted \\
SPURS-GN-2 & 7.19 & Broad [O III], FWHM 1{,}627\cite{tang26} & [O III] $-34\pm18$, FWHM 379 (SPURS); $-149\pm99$, FWHM 697 (FRESCO) & Broad width not recovered \\
SPURS-GN-29 & 7.04 & Broad [O III]; absorber $-140$\cite{tang26} & [O III] $+56\pm24$, FWHM 716; \Ha\ not covered & Recovered; redshifted \\
SPURS-GN-2004 & 6.71 & Broad [O III]; absorber $-293$\cite{tang26} & [O III] $-7\pm11$, FWHM 558; absorber $-103$ & Recovered; at rest; absorber bound \\
RUBIES-55604 (RUBIES-2, CEERS-7902) & 6.98 & Broad [O III]\cite{tang26}; absorber $-305$\cite{juodzbalis26b} & [O III] $+19\pm18$, FWHM 312 (RUBIES); no second component (GO 4287); absorber $-326$ & Broad [O III] not recovered; absorber bound \\
RUBIES-49140 (OCEANS 35829, C3PO 46403) & 6.69 & Absorber $-14$\cite{davis26}; [O III] near rest\cite{wang26} & [O III] $+12\pm11$; absorber $-10$; G395H data exclusive until 2027 & Not blueshifted; absorber bound \\
RUBIES-42046 (OCEANS 20504) & 5.28 & Absorber $-315$\cite{davis26} & [O III] $-78\pm18$ and $-74\pm18$ (G395H), below $\sigma_\star$; absorber $-435$ & Turbulence; absorber bound \\
JADES-GN-68797 & 5.04 & Absorber\cite{juodzbalis26b} & [O III] $-38\pm9$, FWHM 320, below $\sigma_\star$; absorber $-196\pm67$ & Turbulence; absorber bound \\
LRD-204851 (5664-GS-13971) & 5.48 & Blue Ly$\alpha$ peak, $-430$\cite{ji26}; absorber\cite{juodzbalis26b} & No second [O III] component; absorber $-62\pm26$ & Not recovered; absorber bound \\
5664-GN-9771 & 5.54 & Absorber\cite{juodzbalis26b} & [O III] $-22\pm8$, at width bound; absorber $-240$ (IFU) & Absorber bound \\
5664-GN-15498 & 5.09 & Absorber\cite{juodzbalis26b} & No second [O III] component; absorber $-38$ (IFU) & Absorber bound \\
JADES-GN-38147 & 5.87 & Absorber\cite{juodzbalis26b} & [O III] $+77\pm43$, at width bound; absorber $-72\pm21$ & Absorber bound \\
RUBIES-182791 & 4.72 & Absorber $-298$\cite{juodzbalis26b} & No second [O III] component; absorber $-50\pm41$ & Absorber bound \\
UNCOVER-45924 & 4.47 & Broad [O III] near rest\cite{wang26} & [O III] $+14\pm7$, FWHM 368; absorber $+26$ & Recovered; not blueshifted \\
OCEANS 102364 & 4.54 & Absorber $-138$\cite{davis26} & No second [O III] component; absorber $-90$ & Absorber bound \\
OCEANS 100424 & 4.95 & Absorber $+137$\cite{davis26} & No second [O III] component; absorber $+21$ & Absorber bound \\
C3PO 45290 & 8.35 & [O III] $\lambda4364$ outflow disfavoured\cite{papovich26} & [O III] $-230\pm149$, $F$-test $1.9\sigma$ & Not recovered \\
BT-J1148 & 5.01 & Absorber\cite{juodzbalis26b} & Nucleus not located in the public cube & Not reanalysed \\
A2744-QSO1 & 7.04 & Tentative \Ha\ outflow\cite{juodzbalis26} & High-resolution data are IFU only; prism spectrum only in hand & Not reanalysed \\
\bottomrule\end{tabular}\end{table*}

\end{document}